\documentclass[a4paper,onecolumn,11pt]{quantumarticle}
\pdfoutput=1
\usepackage[utf8]{inputenc}
\usepackage[english]{babel}
\usepackage[T1]{fontenc}
\usepackage{amsmath}
\usepackage{hyperref}

\usepackage{geometry}
\usepackage{subcaption}
\usepackage{amsmath}
\usepackage{enumerate}
\usepackage{physics}
\usepackage{xcolor}
\usepackage{mathtools}
\usepackage{hyperref}
\usepackage{bbold}
\usepackage{float}
\usepackage{authblk}
\usepackage{mathrsfs} 

\usepackage[utf8]{inputenc}
\usepackage{pgfplots}
\usepgfplotslibrary{groupplots,dateplot}
\usetikzlibrary{patterns,shapes.arrows}

\usepackage{tikz}
\usepackage{lipsum}
\usepackage{braket}

\newcommand{\comment}[1]{}
\begin{document}

\title{Simulating single-photon experiments with a quantum computer}

\author[1]{Priyasheel Prasad}
\author[2]{Marco Russo}
\author[2]{Bartolomeo Montrucchio}

\affil[1]{priyasheel.rch@gmail.com}
\affil[2]{Department of Control and Computer Engineering (DAUIN), Politecnico di Torino, Turin, Italy}

\maketitle

\begin{abstract}
In this work, we simulate the behavior of photons in a laboratory experiment using a quantum computer and examine how the simulation results compare with the theoretical predictions. The 
experiment involves both protective and non-protective measurements. While the latter involves complete wavefunction collapse, the former combines weak interactions with a $protective$ $mechanism$ thereby preserving the photon wave function coherence until its final detection. The simulation gives insights as to how efficient quantum computers can be in simulating actual physical systems as the amount of computation increases.
\end{abstract}

\section{Background}     
One of the main applications of quantum computing would be the efficient simulation of the dynamics of a quantum system \cite{Feynman1982, Lloyd1996}. This is due to the practical applications they can offer \cite{Lloyd1996} due to the strong evidence of universality in quantum computation \cite{royal, Feynman1982}, as well as the exponential computational advantage they can exhibit, at least in certain cases. In the present work, we simulate a laboratory experiment quite similar to the one mentioned in \cite{Rebufello2021}, involving single photons on a quantum computer. Unlike projective measurements, which involve wavefunction collapse, this experiment involves protective measurement, wherein photons undergo decoherence by weakly interacting with a pair of birefringent crystals and a $protective$ $mechanism$ which involves projections onto the initial photon polarization state. This repeated projections after each weak interaction increases the survival probability of the photon as compared to the case where projection is absent. We first describe the experimental setup and the theoretical results, then the simulation steps, and finally the results.

\section{Experimental setup}

\begin{figure}[t]
    \centering
    \includegraphics[scale=0.09]{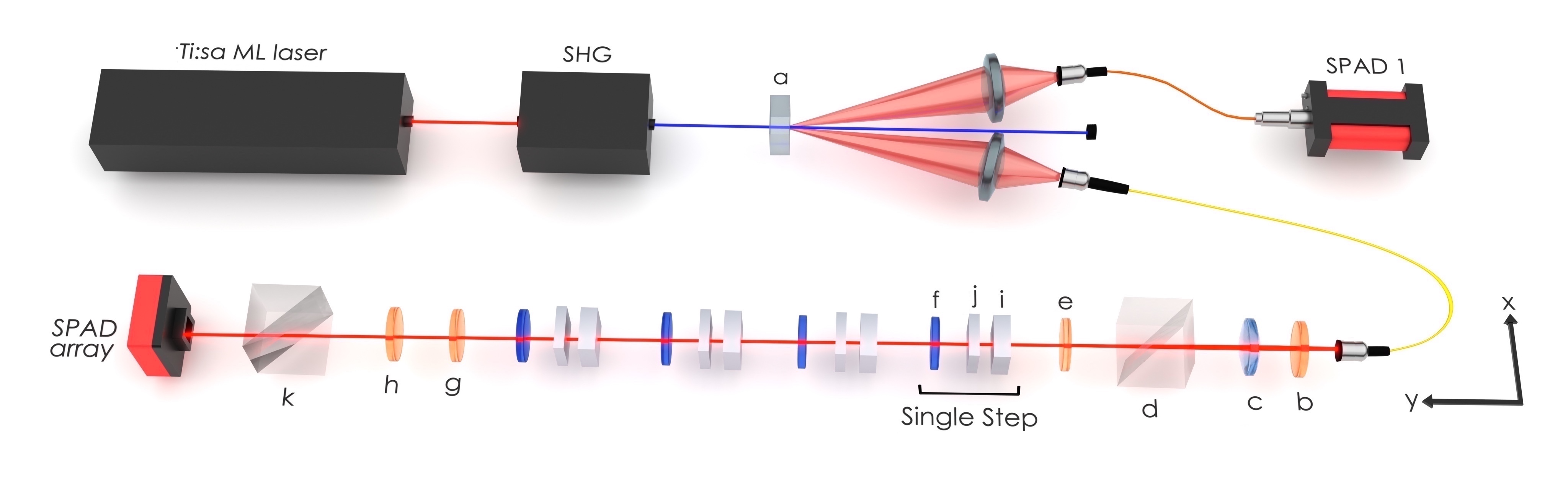}
    \caption{Experimental setup: HWP b is used to adjust the laser beam power. A PBS is placed before b (not shown) to achieve this. The operators for HWP e, g and h are $W_{\pi/8}$, $W_{3\pi/8}$ and $W_{\pi/4}$, respectively (see the appendix). For the $protective$ $case$ the polarizer f in each step is used and for the $non-protective$ $case$, it is removed. The purpose of h is just to swap the reflected and transmitted component of the photon wave function for the PBS k.}
\label{fig:shift}
    \label{fig:setup}
\end{figure}

Figure \ref{fig:setup} shows the experimental setup. Heralded single photons are produced by type I parametric down-conversion in a LiIO$_3$ nonlinear crystal a, producing idler and signal photons. The pump beam is obtained by second-harmonic generation (SHG) of a Ti:Sapphire mode-locked laser output. The idler photons are fiber-coupled and detected by means of a silicon single-photon avalanche diode (SPAD I), sending a trigger pulse to the signal-photon detection system. The signal photons are fiber coupled and the photon beam is collimated as a Gaussian beam that passes through a half-wave plate (HWP) b, a collimating lens c, a polarizing beam splitter (PBS) d, and a HWP e, before entering a sequence of repeating units comprising of a pair of birefringent crystals (i and j; see the appendix for their details) and a polarizer f. The photons then pass through two more HWPs, g and h, thereafter passing through a PBS k before being detected by a SPAD array detector. The horizontally polarized component is transmitted while the vertical one is reflected by both PBS d and k. The polarizer (f) axis is oriented at $45^{\circ}$ with the $x-$axis. In the next section, we deduce the state of a photon before being detected.

\section{Theoretical results}
\noindent
Let the wave function of the photon before HWP e, be $\ket{i} = \ket{H} \ket{\psi} \ket{\tilde{y}}$, where $\ket{H}$ is the linear polarization state of the photon, $\ket{\tilde{y}}$ implies that the photon is propagating in the $y-$direction and $\ket{\psi}$ represents the Gaussian mode of the photon beam, given by 

\begin{equation}
	\ket{\psi} = \int_{-\infty}^{\infty} dx \ket{x}  \braket{x|\psi}
\end{equation} 

where

  \begin{equation}
\braket{x|\psi} = \psi (x) = \frac{1}{(2\pi \sigma^2)^{1/4}}e^{-x^2/4\sigma^2}
  \end{equation}

We first describe the (weak) interaction of the photon with the birefringent crystals. This is modeled using the von Neumann indirect measurement protocol \cite{Aharonov1988, Guth1933, Resch2004}. Assuming that the photon with wave function $\ket{H} \ket{\psi}$  interacted with the crystal i (see the appendix for details) from time $t_0$ to $t$, its wave function after it exits the crystal will be $U(t,t_0) \ket{H}  \ket{\psi}$, where $U(t,t_0) = e^{-\frac{i}{\hbar}\int_{t_0}^{t} H_i(t') dt'}$ with the condition that the interaction Hamiltonian $H_i$ commutes with itself during the interaction, i.e., $[H_i(t),H_i(t')] = 0$. $H_i$ for our case can be taken as \cite{Piacentini2016} $H_i = \gamma \hat{A} \otimes \hat{p}$, $\gamma$ being the coupling constant, $\hat{A} = \ket{H}\bra{H}$ and $\hat{p} = -i\hbar \frac{\partial}{\partial x}$ being the momentum operator. As $H_i$ does not depend on time, therefore $U = e^{-\frac{i}{\hbar}g\hat{A}\otimes \hat{p}}$, where $g = \gamma (t-t_0)$ is the coupling strength. The wave function of the photon after it exits the birefringent crystal will be
\begin{widetext}
\begin{align}
\label{eq:birefr}
	U \ket{H} \otimes \ket{\psi} &= e^{-\frac{i}{\hbar}g \ket{H} \bra{H} \otimes \hat{p}} \ket{H} \otimes \ket{\psi} \nonumber \\
	&= \bigg[ \mathbb{1} -\frac{i}{\hbar}g \ket{H}\bra{H} \otimes \hat{p} + \frac{i^2}{2! \hbar^2}g^2 \ket{H}\bra{H} \otimes \hat{p}^2 - ...\bigg] \ket{H} \otimes \ket{\psi} \nonumber \\
	&= \bigg[ \ket{H} -\frac{i}{\hbar}g \ket{H} \otimes \hat{p} + \frac{i^2}{2! \hbar^2}g^2 \ket{H} \otimes \hat{p}^2 - ...\bigg] \ket{\psi} \nonumber \\
	&= \ket{H} \otimes e^{-\frac{i}{\hbar}g \hat{p}} \ket{\psi} \nonumber \\
	&= \ket{H} \otimes  \ket{\psi}_{x-g}
\end{align} 
\end{widetext}

In the last step, $e^{-\frac{i}{\hbar}g\hat{p}}$ acts as a translation operator \cite{sakurai2014modern} on the wave function $\ket{\psi}$. Its effect is to transform the $\psi(x)$ amplitudes into $\psi(x-g)$, so that \begin{equation*}
    \ket{\psi}_{x-g} = \int_{-\infty}^{\infty} dx \;\psi(x-g)\ket{x} 
\end{equation*}

Crystal j is used only for phase compensation, as the $o-$ray and $e-$ray in crystal i follow
different paths. The vertical polarization state $\ket{V}$ would pass through the birefringent crystals unchanged.
\\
\\
When the polarizer f is used in each step, we call it $protective$ $case$ and if it is not used in each step, we call it $non-protective$ $case$. If the fast axis of HWPs e, g and h are inclined at $\alpha$, $\beta = 3\pi/8$ and $\pi/4$, the wave function of the exiting photon (after k) for the protective case can be obtained as

\begin{equation}
    \ket{f} \propto \frac{1}{2^{n- 1/2}} \Bigg[\sum_{k=0}^{n} \ket{\psi}_{x -kg} \Bigg\{ {{n-1}\choose k} \sin 2\alpha   + {{n-1}\choose {k-1}} \cos 2\alpha \Bigg\} \Bigg] \ket{H} \ket{\tilde{y}}
\end{equation}

where $n$ is the number of steps and $ {n\choose k} = 0$ if $k<0$ or $k>n$. The survival probability of the exiting photon (detected by the SPAD array) will be

\begin{equation}
	 P(n) = \frac{1}{2^{2n -1}} \int_{-\infty}^{\infty} \Bigg[ \sum_{k=0}^{n} \psi(x-kg) \Bigg\{ {{n-1}\choose k} \sin 2\alpha   + {{n-1}\choose {k-1}} \cos 2\alpha \Bigg\} \Bigg]^2 dx 
\end{equation} 

For $non-protective$ $case$, the wave function of the exiting photon will be

    \begin{equation}
    \begin{split}
        \ket{f'} &= \frac{\sin 2\alpha \ket{\psi}_x + \cos 2\alpha \ket{\psi}_{x-ng}}{\sqrt 2} \ket{H} \ket{\tilde{y}} \\ 
        &+ \frac{\sin 2\alpha  \ket{\psi}_{y-\ell} - \cos 2\alpha \ket{\psi}_{y-\ell+ng}}{\sqrt 2}\ket{V}\ket{-\tilde{x}}
            \end{split}
    \end{equation}

where $\ell$ is the linear distance traveled by the photon beam, depending on where the origin is chosen along the beam. The probabilities of detecting a photon transmitted and reflected by PBS k will, respectively, be

\begin{align}
	P'_{T}(n) = \frac{1}{2}\int_{-\infty}^{\infty} \big[\sin 2\alpha \psi(x) + \cos 2\alpha  \psi(x - ng) \big]^2   dx
\end{align}

\begin{align}
	P'_{R}(n) = \frac{1}{2} \int_{-\infty}^{\infty} \big[ \sin 2\alpha \psi(x) - \cos 2\alpha \psi(x +ng) \big]^2   dx
\end{align}

The plot of survival probabilities vs. $n$ is shown in figure 2. $P'_{T}(n) + P'_{R}(n) = 1$ as the photon is ought to be found at one of the ports.

\begin{figure}[H]
\centering
\begin{tikzpicture}
\definecolor{darkgray176}{RGB}{176,176,176}
\definecolor{darkviolet1910191}{RGB}{191,0,191}
\definecolor{green01270}{RGB}{0,127,0}
\definecolor{lightgray204}{RGB}{204,204,204}

\begin{axis}[
legend cell align={left},
legend style={
  fill opacity=0.8,
  draw opacity=1,
  text opacity=1,
  at={(0.97,0.03)},
  anchor=south east,
  draw=lightgray204
},
tick align=outside,
tick pos=left,
title={},
x grid style={darkgray176},
xlabel={Number of steps ($n$)},
xmin=-1.5, xmax=31.5,
xtick style={color=black},
y grid style={darkgray176},
ylabel={Probability},
ymin=-0.0500000000000003, ymax=1.05000000000001,
ytick style={color=black}
]
\addplot [semithick, blue, dashed, mark=*, mark size=1, mark options={solid}]
table {%
0 1
1 0.995630145126226
2 0.982748395879233
3 0.962018518412417
4 0.934483556101105
5 0.901478897728775
6 0.864525168490726
7 0.825212876792357
8 0.785090906168803
9 0.745569492762732
10 0.707845621212714
11 0.672855325081405
12 0.64125377543634
13 0.613420820124419
14 0.589487200011542
15 0.569375235969446
16 0.552847377970744
17 0.539556497776407
18 0.529092941901619
19 0.521024847938566
20 0.514929783320556
21 0.51041716549015
22 0.50714201963517
23 0.504811360890025
24 0.503184859256043
25 0.502071514481673
26 0.501323918101094
27 0.500831399356152
28 0.500513019254397
29 0.500311052047165
30 0.500185313850602
};
\addlegendentry{\begin{footnotesize}$P'_{T}(n)$\end{footnotesize}}
\addplot [semithick, green01270, dashed, mark=*, mark size=1, mark options={solid}]
table {%
0 0
1 0.00436985487377359
2 0.0172516041207599
3 0.0379814815875826
4 0.0655164438989019
5 0.0985211022712241
6 0.135474831509272
7 0.174787123207646
8 0.214909093831197
9 0.254430507237268
10 0.292154378787286
11 0.327144674918595
12 0.35874622456366
13 0.386579179875581
14 0.410512799988449
15 0.430624764030554
16 0.447152622029256
17 0.460443502223593
18 0.470907058098382
19 0.478975152061434
20 0.485070216679433
21 0.489582834509831
22 0.492857980364812
23 0.495188639109951
24 0.496815140743957
25 0.497928485518327
26 0.498676081898905
27 0.499168600643848
28 0.499486980745603
29 0.499688947952835
30 0.499814686149398
};
\addlegendentry{\begin{footnotesize}$P'_{R}(n)$\end{footnotesize}}
\addplot [semithick, cyan, dashed, mark=*, mark size=1, mark options={solid}]
table {%
0 1
1 0.999999999999999
2 0.999999999999993
3 1
4 1.00000000000001
5 0.999999999999999
6 0.999999999999998
7 1
8 1
9 1
10 1
11 1
12 1
13 1
14 0.999999999999991
15 1
16 1
17 1
18 1
19 1
20 0.999999999999989
21 0.999999999999982
22 0.999999999999982
23 0.999999999999977
24 1
25 1
26 1
27 1
28 1
29 1
30 1
};
\addlegendentry{\begin{footnotesize}$P'_{T}(n) + P'_{R}(n)$\end{footnotesize}}
\addplot [semithick, red, dashed, mark=*, mark size=1, mark options={solid}]
table {%
0 1
1 0.995630145126226
2 0.991317244096027
3 0.987060066911328
4 0.982857420548247
5 0.978708147536386
6 0.974611124604331
7 0.970565261387916
8 0.966569499197769
9 0.962622809842958
10 0.958724194507729
11 0.954872682678529
12 0.951067331118643
13 0.947307222887633
14 0.943591466405919
15 0.939919194554026
16 0.936289563818358
17 0.932701753467578
18 0.929154964766185
19 0.925648420220489
20 0.922181362856606
21 0.918753055527495
22 0.915362780249026
23 0.912009837562855
24 0.908693545925086
25 0.90541324111945
26 0.902168275693876
27 0.898958018419317
28 0.89578185376983
29 0.892639181422908
30 0.889529415779149
};
\addlegendentry{\begin{footnotesize}$P(n)$ \end{footnotesize}}
\end{axis}

\end{tikzpicture}
\caption{The standard deviation of the Gaussian distribution is $\sigma = 0.4$ and the coupling strength is taken as $g =$ 106 $\mu m$. These plots are for $\alpha = \pi/8$ and $\beta = 3\pi/8$, the same values which we take for the simulation. It can be shown that $\displaystyle{\lim_{n\to\infty} P(n) = 0 }$ and $\displaystyle{\lim_{n\to\infty} P'_T(n) = \lim_{n\to\infty} P'_R(n) = 1/2}$. The $protective$ $mechanism$ works only for a limited number of steps, as for some $n$, $P(n)$ would definitely become less than $P'_T(n)$.}
\end{figure}
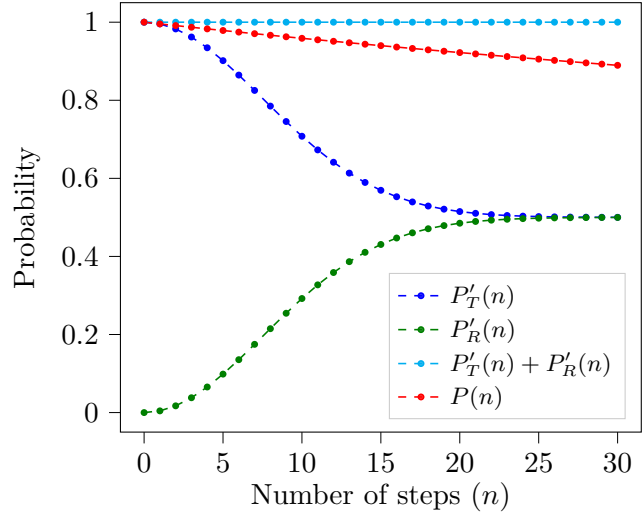

\section{Simulating the experiment}

Let most of the power of the laser beam be confined in a radius of 1.5 $mm$. We
consider $\ket{\psi}$ to be confined in a finite region $-d \leq x \leq d$, where $d = 3$ $mm$. This is a valid assumption as a Gaussian distribution is mostly confined in 6$\sigma$ and we take $\sigma = 0.4$ for the simulation. If we represent the total wave function of the photon using seven qubits (one for the photon polarization state and $n=6$ for $\ket{\psi}$), then the differential step size $\Delta x = \frac{2d}{2^n - 1} = 95.24$ $\mu m$ is less than $g =$ 106 $\mu m$. Seven qubits are therefore sufficient to represent the photon wave function. $\ket{\psi}$ can be written as 

\begin{equation} \label{eq:10}
	\ket{\psi} = \sum_{j=0}^{N -1} a_j \ket{j}
\end{equation} 

where we consider the discretized version of the normal distribution on $N = 2^n$ equidistant points $a_j$ and $\ket{0}$, $\ket{1}$, ..., $\ket{j}$, ..., $\ket{N -1}$ are the computational basis states of the $n$ qubits. We next discuss how to simulate the effect of the birefringent crystals and the polarizer.

\subsection{Simulating the effect of birefringent crystals}
We use the relation $\hat{p}\psi(x) = \mathcal{F}^{-1} p \mathcal{F}\psi(x) = \mathcal{F}^{-1}p \phi(p)$ where $\phi(p)$ is the Fourier transform ($\mathcal{F}$) of $\psi(x)$, to switch to the momentum basis where $\hat{p}$ is diagonal and then back to the position basis. In order to simulate the effect of birefringent crystals, we start with \ref{eq:birefr} and use the inverse quantum Fourier transform $U^{-1}_{\textnormal{QFT}}$ instead of the continuous $\mathcal{F}$ and vice versa (see the appendix), with the property that $U_{\textnormal{QFT}}$ is unitary. We have

\begin{widetext}
    \begin{align} \label{eq:wf_qft}
	\ket{H} \otimes e^{-\frac{i}{\hbar}g \hat{p}} \ket{\psi} &= \ket{H} \otimes e^{-\frac{i}{\hbar}g U_{\textnormal{QFT}} \hat{p} U^{-1}_{\textnormal{QFT}}} \ket{\psi} \nonumber \\ 
	&= \ket{H} \otimes \bigg[ \mathbb{1} - \frac{i}{\hbar}g U_{\textnormal{QFT}} \hat{p} U^{-1}_{\textnormal{QFT}} + \frac{i^2}{2!\hbar^2}g^2 U_{\textnormal{QFT}} \hat{p}^2 U^{-1}_{\textnormal{QFT}} - ... \bigg] \ket{\psi} \nonumber \\ 
	&= \ket{H} \otimes U_{\textnormal{QFT}} \bigg[ \mathbb{1} - \frac{i}{\hbar}g\hat{p} + \frac{i^2}{2!\hbar^2}g^2\hat{p}^2 - ... \bigg] U^{-1}_{\textnormal{QFT}}  \ket{\psi} \nonumber \\
	&= \ket{H} \otimes U_{\textnormal{QFT}} e^{-\frac{i}{\hbar}g \hat{p}} U^{-1}_{\textnormal{QFT}} \ket{\psi} \nonumber \\
	&= \ket{H} \otimes U_{\textnormal{QFT}} e^{-\frac{i}{\hbar}g {p}} \ket{\phi}
    \end{align} 
\end{widetext}

We can apply $U^{-1}_{\textnormal{QFT}}$ on $\ket{\psi}$ to obtain $\ket{\phi}$, the wave function in the momentum space, then apply $e^{-\frac{i}{\hbar} g p}$ and finally apply $U_{\textnormal{QFT}}$ to obtain the wave function again in position space. The transformation $\ket{p} \to e^{i\lambda f(p)} \ket{p}$ ($\lambda$ is a real number), where $f(p) = -\frac{g}{\hbar}p$ and $p = $ 0, 1, 2, ..., $N-1$, can be achieved by applying controlled phase shifts \cite{Benenti_2008} as shown in figure \ref{fig:shift}. Using only the postulates of quantum mechanics that form the basis of quantum computation, there is no way to determine the actual numerical value of the constant $\hbar$. We take $\hbar = 1$ for the simulation. Consequently, we are forced to introduce the factor $\lambda$ which should have a fixed value for the circuit shown in figure \ref{fig:shift} and the way it is determined is discussed in the last section.

\begin{figure}[H]
\centering  
\input{shift0.tex}
\caption{Quantum circuit to perform the transformation $\ket{p} \to e^{i\lambda f(p)}\ket{p}$ where $f(p)=-\frac{g}{\hbar}p$, $p \in \{0, 1, 2, ..., 2^n-1\}$ where $n=6$ is the number of qubits representing the Gaussian state $\ket{\psi}$. The gate $F_j = \footnotesize{\begin{bmatrix} e^{if(2j)} & 0 \\ 0 & e^{if(2j+1)} \end{bmatrix}}$, where $j \in \{0, 1, 2, ..., 2^{n'} -1\}$ with $n' = n-1$, is applied to the $n^{\textnormal{th}}$ qubit and the first $n'$ qubits act as control qubits. If $j=\sum_{i=0}^{n'-1} j_i 2^i$ where $j_i \in \{0,1\}$, then for $F_j$, the conditional on $i^{\textnormal{th}}$ qubit where $i \in \{0, 1, 2, ..., n'-1\}$, is set to zero or one depending on whether $j_{i}$ is 0 or 1, respectively. The circuit implements the transformation that can be represented by a $2^n \times 2^n$ unitary matrix $U_p = \textnormal{diag}(F_0, ..., F_j, ..., F_{31})$.}
\label{fig:shift}
\end{figure}

The transformation $\ket{p} \to e^{i\lambda f(p)} \ket{p}$ can also be achieved by evaluating the function $f(p)$ \cite{nielsen2010chapter4}. Assuming that the function $f:\{0,...,2^m-1\} \to \{0,...,2^n-1\}$, where $m$ and $n$ are positive integers, can be computed reversibly using the ancillary register $\ket{y}$, i.e., the operation $\ket{p,y} \to \ket{p,y\oplus f(p)}$ can be performed reversibly, followed by the transformation $\ket{p,y\oplus f(p)} \to e^{i\lambda f(p)} \ket{p,y\oplus f(p)}$ and finally the uncomputation step $e^{if(p)} \ket{p,y\oplus f(p)} \to e^{i\lambda f(p)} \ket{p,y}$, thereby implementing the desired transformation. Here, $\oplus$ represents the addition modulo $2^n$ operation. If $f(p) = p$ then we can skip the first and third steps and invoke just the second step by taking only the qubits that represent the pointer wave function, i.e., without using any ancillary qubit. We have for $p \in [0, 2^n - 1]$ and $p=\sum_{j=0}^{n-1} p_j 2^j$ where $p_j \in \{0,1\}$, $e^{i\lambda p} = e^{\sum_{j=0}^{n-1} i\lambda p_j 2^j} = \prod_{j=0}^{n-1} e^{i\lambda p_j2^j}$, which can be implemented using $n$ single qubit gates. The matrix representation of such a gate that acts on the $j^{th}$ qubit is given by $U_{j} = \begin{bmatrix} 1 & 0\\0 & e^{i\lambda 2^j}\end{bmatrix}$. We would call the transformation $\ket{p} \to e^{i\lambda f(p)} \ket{p}$ as $\textit{phase shift}$. The operation $U^{-1}_{\textnormal{QFT}}-\textit{phase shift}-U_{\textnormal{QFT}}$ implements the translation of the pointer state $\ket{\psi}$.

\subsection{Simulating the effect of polarizer}
\noindent
The operator for the polarizer f is the projector $P_{\pi/4} = \ket{+} \bra{+}$, which is non-unitary. To simulate this \cite{Childs2012}, we consider $P_{\pi/4}$ as a linear combination of unitary operators,

\begin{align}
	P_{\pi/4} = \frac{1}{2} (2P_{\pi/4} - I) + \frac{I}{2} = \frac{X}{2} + \frac{I}{2}
\end{align}

Let $\ket{\phi}$ be the wave function of the system on which $P_{\pi/4}$ is to be applied. Consider the following circuit where the first qubit is an ancilla qubit. 

\begin{figure}[H]
\centering
\begin{tikzpicture}
\draw [line width=0.5pt] (-2.7,0.2)-- (2,0.2);
\draw[color=black] (-3,0.2) node {$\ket{0}$};

\draw [line width=0.5pt] (-2.7,-1)-- (2,-1);
\draw[color=black] (-3.03,-1) node {$\ket{\phi}$};

\fill[line width=0.5pt,color=black,fill=white,fill opacity=1.0] (-2.1,-0.05) -- (-1.6,-0.05) -- (-1.6,0.45) -- (-2.1,0.45) -- cycle;
\draw [line width=0.5pt,color=black] (-2.1,-0.05)-- (-1.6,-0.05);
\draw [line width=0.5pt,color=black] (-1.6,-0.05)-- (-1.6,0.45);
\draw [line width=0.5pt,color=black] (-1.6,0.45)-- (-2.1,0.45);
\draw [line width=0.5pt,color=black] (-2.1,0.45)-- (-2.1,-0.05);
\begin{scriptsize}
\draw[color=black] (-1.85,0.2) node {$H$};
\end{scriptsize}

\draw [line width=0.5pt] (-1,0.2)-- (-1,-1);
\draw [fill=black, line width=0.5] (-1,0.2) circle (1.6pt);

\fill[line width=0.5pt,color=black,fill=white,fill opacity=1.0] (-0.4,-0.05) -- (0.1,-0.05) -- (0.1,0.45) -- (-0.4,0.45) -- cycle;
\draw [line width=0.5pt,color=black] (-0.4,-0.05)-- (0.1,-0.05);
\draw [line width=0.5pt,color=black] (0.1,-0.05)-- (0.1,0.45);
\draw [line width=0.5pt,color=black] (0.1,0.45)-- (-0.4,0.45);
\draw [line width=0.5pt,color=black] (-0.4,0.45)-- (-0.4,-0.05);
\begin{scriptsize}
\draw[color=black] (-0.15,0.2) node {$H$};
\end{scriptsize}

\fill[line width=0.5pt,color=black,fill=white,fill opacity=1.0] (0.7,-0.05) -- (1.4,-0.05) -- (1.4,0.45) -- (0.7,0.45) -- cycle;
\draw [line width=0.5pt,color=black] (0.7,-0.05)-- (1.4,-0.05);
\draw [line width=0.5pt,color=black] (1.4,-0.05)-- (1.4,0.45);
\draw [line width=0.5pt,color=black] (1.4,0.45)-- (0.7,0.45);
\draw [line width=0.5pt,color=black] (0.7,0.45)-- (0.7,-0.05);
\draw [shift={(1.05,-0.16)},line width=0.4pt]  plot[domain=40:140,variable=\x]({0.38*cos(\x)}, {0.38*sin(\x)});
\draw [line width=0.4pt] (1.05,0.06)-- (1.27,0.38);

\fill[line width=0.5pt,color=black,fill=white,fill opacity=1.0] (-1.25,-1.25) -- (-0.75,-1.25) -- (-0.75,-0.75) -- (-1.25,-0.75) -- cycle;
\draw [line width=0.5pt,color=black] (-1.25,-1.25)-- (-0.75,-1.25);
\draw [line width=0.5pt,color=black] (-0.75,-1.25)-- (-0.75,-0.75);
\draw [line width=0.5pt,color=black] (-0.75,-0.75)-- (-1.25,-0.75);
\draw [line width=0.5pt,color=black] (-1.25,-0.75)-- (-1.25,-1.25);
\begin{scriptsize}
\draw[color=black] (-1,-1.0) node {$X$};
\end{scriptsize}

\end{tikzpicture}
\caption{}
\end{figure}

The state of the system will evolve as
\begin{align*}
	\ket{0} \ket{\phi} &\xrightarrow[\text{}]{\text{$H$}} \frac{1}{\sqrt 2}\big[ \ket{0} + \ket{1} \big] \ket{\phi} \\
	&\xrightarrow[\text{}]{\text{$CX$}} \frac{1}{\sqrt 2}\big[ \ket{0} \ket{\phi} + \ket{1} X \ket{\phi} \big] \\
	&\xrightarrow[\text{}]{\text{$H$}} \ket{0} \bigg[ \frac{I}{2} + \frac{X}{2} \bigg] \ket{\phi} +  \ket{1} \bigg[ \frac{I}{2} - \frac{X}{2} \bigg] \ket{\phi}
\end{align*}  

If the ancilla qubit is measured and found to be 0, then only $\frac{I}{2} + \frac{X}{2}$ acts on $\ket{\phi}$. The readings for which the ancilla measures 1 are discarded. We now describe the steps to simulate the experiment.

\subsection{Simulation steps}
\begin{figure}[H]
\centering  \input{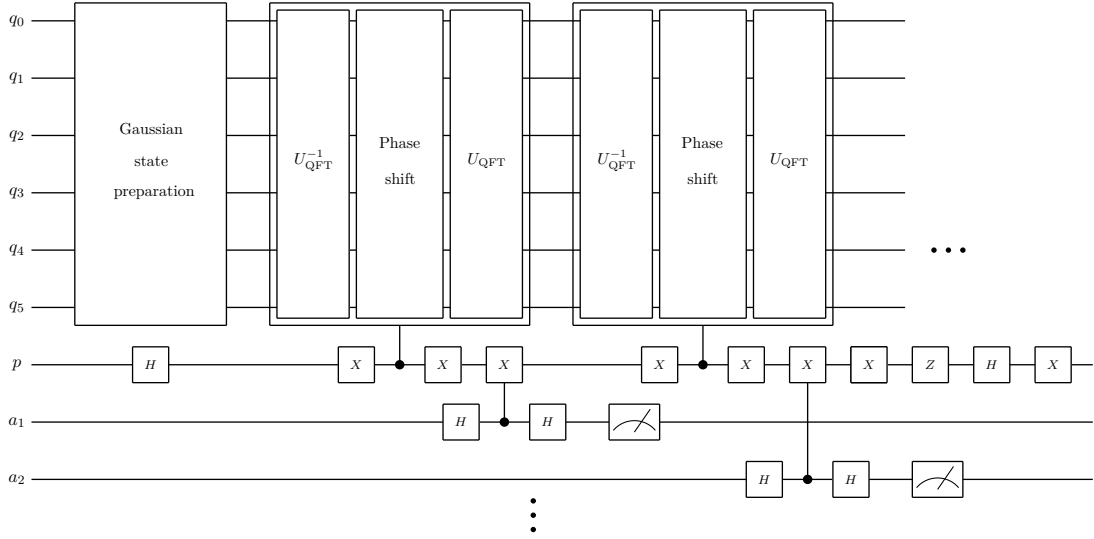}
\caption{Quantum circuit to simulate the protective case: qubit $p$ represents the polarization state of the photon and $H$, $Z$ and $X$ are the Hadamard gate and the Pauli $Z$ and $X$ gate, respectively. The operators for HWPs e and g are $H$ and $HZX$, respectively. The phase shift operation is performed by the circuit shown in figure \ref{fig:shift}, acting on pointer state $\ket{\psi}$. $U_{\textnormal{QFT}}$ and $U^{-1}_{\textnormal{QFT}}$ represent the quantum Fourier transform and its inverse, respectively.}
\label{fig:circuit_protective}
\end{figure}

\noindent
For the protective case, $n$ steps can be simulated using the circuit shown in figure \ref{fig:circuit_protective}. Here, $n$ ancilla qubits $a_1$, $a_2$, ..., $a_n$ are used and $p$ represents the polarization state of the photon. $\ket{0}$ and $\ket{1}$ represent horizontal and vertical polarization states, respectively.

 Starting with the initial state $\ket{\psi} \ket{0}_p \otimes_{i=1}^{n} \ket{0}_{a_i}$, the state of the quantum computer $\ket{\alpha}$ after the $n$ ($n>0$) steps, assuming that the ancilla qubits $a_1$, $a_2$, ..., $a_{n-1}$ were measured and found to be 0, will be

  \begin{align}
	 \ket{\alpha} & \propto \frac{1}{2^{n}} \Bigg[ \sum_{k=0}^{n} {n\choose k} \ket{\psi}_{x-kg}  \Bigg] \ket{0}_p \ket{0}_{a_n}  \nonumber  + \text{terms of } \ket{1}_{a_n}
\end{align}

\vspace{0.3cm}

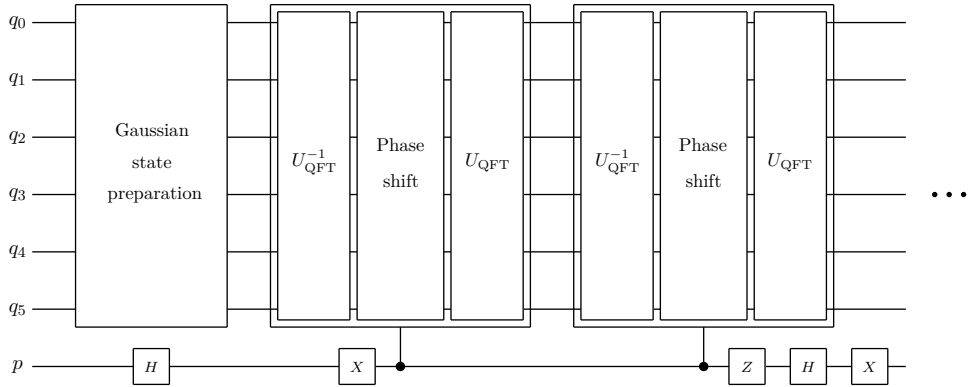
\begin{figure}[H]

\centering
    \begin{tikzpicture}[thick,scale=0.95, every node/.style={scale=0.7}]
\draw [line width=0.5pt] (-2.7,3.8)-- (9.4,3.8);
\draw[color=black] (-2.9,3.8) node {$q_0$};

\draw [line width=0.5pt] (-2.7,3.)-- (9.4,3.);
\draw[color=black] (-2.9,3.) node {$q_1$};

\draw [line width=0.5pt] (-2.7,2.2)-- (9.4,2.2);
\draw[color=black] (-2.9,2.2) node {$q_2$};

\draw [line width=0.5pt] (-2.7,1.4)-- (9.4,1.4);
\draw[color=black] (-2.9,1.4) node {$q_3$};

\draw [line width=0.5pt] (-2.7,0.6)-- (9.4,0.6);
\draw[color=black] (-2.9,0.6) node {$q_4$};

\draw [line width=0.5pt] (-2.7,-0.2)-- (9.4,-0.2);
\draw[color=black] (-2.9,-0.2) node {$q_5$};

\draw [line width=0.5pt] (-2.7,-1)-- (9.4,-1);
\draw[color=black] (-2.9,-1) node {$p$};

\fill[line width=0.5pt,color=black,fill=white,fill opacity=1.0] (-2.1,-0.45) -- (0,-0.45) -- (0,4.05) -- (-2.1,4.05) -- cycle;
\draw [line width=0.5pt,color=black] (-2.1,4.05)-- (-2.1,-0.45);
\draw [line width=0.5pt,color=black] (-2.1,4.05)-- (0,4.05);
\draw [line width=0.5pt,color=black] (0,-0.45)-- (-2.1,-0.45);
\draw [line width=0.5pt,color=black] (0,-0.45)-- (0,4.05);
\begin{small}
\draw[color=black] (-1.04,2.3) node {Gaussian};
\draw[color=black] (-1.04,1.85) node {state};
\draw[color=black] (-1.0,1.4) node {preparation};
\end{small}

\draw [line width=0.5pt,color=black] (4.2,4.05)-- (4.2,-0.45);
\draw [line width=0.5pt,color=black] (4.2,4.05)-- (0.6,4.05);
\draw [line width=0.5pt,color=black] (0.6,-0.45)-- (4.2,-0.45);
\draw [line width=0.5pt,color=black] (0.6,-0.45)-- (0.6,4.05);
\fill[line width=0.5pt,color=black,fill=white,fill opacity=1.0] (0.7,-0.35) -- (1.7,-0.35) -- (1.7,3.95) -- (0.7,3.95) -- cycle;
\draw [line width=0.5pt,color=black] (1.7,3.95)-- (1.7,-0.35); 
\draw [line width=0.5pt,color=black] (1.7,3.95)-- (0.7,3.95); 
\draw [line width=0.5pt,color=black] (0.7,-0.35)-- (1.7,-0.35); 
\draw [line width=0.5pt,color=black] (0.7,-0.35)-- (0.7,3.95); 
\begin{small}
\draw[color=black] (1.21,1.85) node {$U^{-1}_{\textnormal{QFT}}$};
\end{small}
\fill[line width=0.5pt,color=black,fill=white,fill opacity=1.0] (1.8,-0.35) -- (3,-0.35) -- (3,3.95) -- (1.8,3.95) -- cycle;
\draw [line width=0.5pt,color=black] (3,3.95)-- (3,-0.35); 
\draw [line width=0.5pt,color=black] (3,3.95)-- (1.8,3.95); 
\draw [line width=0.5pt,color=black] (1.8,-0.35)-- (3,-0.35); 
\draw [line width=0.5pt,color=black] (1.8,-0.35)-- (1.8,3.95); 
\begin{small}
\draw[color=black] (2.4,2.1) node {Phase};
\draw[color=black] (2.41,1.6) node {shift};
\end{small}
\fill[line width=0.5pt,color=black,fill=white,fill opacity=1.0] (3.1,-0.35) -- (4.1,-0.35) -- (4.1,3.95) -- (3.1,3.95) -- cycle;
\draw [line width=0.5pt,color=black] (4.1,3.95)-- (4.1,-0.35); 
\draw [line width=0.5pt,color=black] (4.1,3.95)-- (3.1,3.95); 
\draw [line width=0.5pt,color=black] (3.1,-0.35)-- (4.1,-0.35); 
\draw [line width=0.5pt,color=black] (3.1,-0.35)-- (3.1,3.95); 
\begin{small}
\draw[color=black] (3.61,1.85) node {$U_{\textnormal{QFT}}$};
\end{small}

\draw [line width=0.5pt,color=black] (8.4,4.05)-- (8.4,-0.45);
\draw [line width=0.5pt,color=black] (8.4,4.05)-- (4.8,4.05);
\draw [line width=0.5pt,color=black] (4.8,-0.45)-- (8.4,-0.45);
\draw [line width=0.5pt,color=black] (4.8,-0.45)-- (4.8,4.05);
\fill[line width=0.5pt,color=black,fill=white,fill opacity=1.0] (4.9,-0.35) -- (5.9,-0.35) -- (5.9,3.95) -- (4.9,3.95) -- cycle;
\draw [line width=0.5pt,color=black] (5.9,3.95)-- (5.9,-0.35); 
\draw [line width=0.5pt,color=black] (5.9,3.95)-- (4.9,3.95); 
\draw [line width=0.5pt,color=black] (4.9,-0.35)-- (5.9,-0.35); 
\draw [line width=0.5pt,color=black] (4.9,-0.35)-- (4.9,3.95); 
\begin{small}
\draw[color=black] (5.41,1.85) node {$U^{-1}_{\textnormal{QFT}}$};
\end{small}
\fill[line width=0.5pt,color=black,fill=white,fill opacity=1.0] (6,-0.35) -- (7.2,-0.35) -- (7.2,3.95) -- (6,3.95) -- cycle;
\draw [line width=0.5pt,color=black] (7.2,3.95)-- (7.2,-0.35); 
\draw [line width=0.5pt,color=black] (7.2,3.95)-- (6,3.95); 
\draw [line width=0.5pt,color=black] (6,-0.35)-- (7.2,-0.35); 
\draw [line width=0.5pt,color=black] (6,-0.35)-- (6,3.95); 
\begin{small}
\draw[color=black] (6.59,2.1) node {Phase};
\draw[color=black] (6.6,1.6) node {shift};
\end{small}
\fill[line width=0.5pt,color=black,fill=white,fill opacity=1.0] (7.3,-0.35) -- (8.3,-0.35) -- (8.3,3.95) -- (7.3,3.95) -- cycle;
\draw [line width=0.5pt,color=black] (8.3,3.95)-- (8.3,-0.35); 
\draw [line width=0.5pt,color=black] (8.3,3.95)-- (7.3,3.95); 
\draw [line width=0.5pt,color=black] (7.3,-0.35)-- (8.3,-0.35); 
\draw [line width=0.5pt,color=black] (7.3,-0.35)-- (7.3,3.95); 
\begin{small}
\draw[color=black] (7.8,1.85) node {$U_{\textnormal{QFT}}$};
\end{small}

\draw [line width=0.5pt] (2.4,-0.45)-- (2.4,-1);
\draw [fill=black,line width=0.5] (2.4,-1) circle (1.6pt);

\draw [line width=0.5pt] (6.6,-0.45)-- (6.6,-1);
\draw [fill=black,line width=0.5] (6.6,-1) circle (1.6pt);

\draw [fill=black,line width=0.5] (9.8,1.4) circle (0.8pt);
\draw [fill=black,line width=0.5] (10,1.4) circle (0.8pt);
\draw [fill=black,line width=0.5] (10.2,1.4) circle (0.8pt);

\fill[line width=0.5pt,color=black,fill=white,fill opacity=1.0] (-1.3,-1.25) -- (-0.8,-1.25) -- (-0.8,-0.75) -- (-1.3,-0.75) -- cycle;
\draw [line width=0.5pt,color=black] (-1.3,-1.25)-- (-0.8,-1.25);
\draw [line width=0.5pt,color=black] (-0.8,-1.25)-- (-0.8,-0.75);
\draw [line width=0.5pt,color=black] (-0.8,-0.75)-- (-1.3,-0.75);
\draw [line width=0.5pt,color=black] (-1.3,-0.75)-- (-1.3,-1.25);
\begin{scriptsize}
\draw[color=black] (-1.05,-1.0) node {$H$};
\end{scriptsize}

\fill[line width=0.5pt,color=black,fill=white,fill opacity=1.0] (1.55,-1.25) -- (2.05,-1.25) -- (2.05,-0.75) -- (1.55,-0.75) -- cycle;
\draw [line width=0.5pt,color=black] (1.55,-1.25)-- (2.05,-1.25);
\draw [line width=0.5pt,color=black] (2.05,-1.25)-- (2.05,-0.75);
\draw [line width=0.5pt,color=black] (2.05,-0.75)-- (1.55,-0.75);
\draw [line width=0.5pt,color=black] (1.55,-0.75)-- (1.55,-1.25);
\begin{scriptsize}
\draw[color=black] (1.8,-1.0) node {$X$};
\end{scriptsize}

\fill[line width=0.5pt,color=black,fill=white,fill opacity=1.0] (6.95,-1.25) -- (7.45,-1.25) -- (7.45,-0.75) -- (6.95,-0.75) -- cycle;
\draw [line width=0.5pt,color=black] (6.95,-1.25)-- (7.45,-1.25);
\draw [line width=0.5pt,color=black] (7.45,-1.25)-- (7.45,-0.75);
\draw [line width=0.5pt,color=black] (7.45,-0.75)-- (6.95,-0.75);
\draw [line width=0.5pt,color=black] (6.95,-0.75)-- (6.95,-1.25);
\begin{scriptsize}
\draw[color=black] (7.2,-1.0) node {$Z$};
\end{scriptsize}

\fill[line width=0.5pt,color=black,fill=white,fill opacity=1.0] (7.8,-1.25) -- (8.3,-1.25) -- (8.3,-0.75) -- (7.8,-0.75) -- cycle;
\draw [line width=0.5pt,color=black] (7.8,-1.25)-- (8.3,-1.25);
\draw [line width=0.5pt,color=black] (8.3,-1.25)-- (8.3,-0.75);
\draw [line width=0.5pt,color=black] (8.3,-0.75)-- (7.8,-0.75);
\draw [line width=0.5pt,color=black] (7.8,-0.75)-- (7.8,-1.25);
\begin{scriptsize}
\draw[color=black] (8.05,-1.0) node {$H$};
\end{scriptsize}

\fill[line width=0.5pt,color=black,fill=white,fill opacity=1.0] (8.65,-1.25) -- (9.15,-1.25) -- (9.15,-0.75) -- (8.65,-0.75) -- cycle;
\draw [line width=0.5pt,color=black] (8.65,-1.25)-- (9.15,-1.25);
\draw [line width=0.5pt,color=black] (9.15,-1.25)-- (9.15,-0.75);
\draw [line width=0.5pt,color=black] (9.15,-0.75)-- (8.65,-0.75);
\draw [line width=0.5pt,color=black] (8.65,-0.75)-- (8.65,-1.25);
\begin{scriptsize}
\draw[color=black] (8.9,-1.0) node {$X$};
\end{scriptsize}

\end{tikzpicture}
\caption{Quantum circuit to simulate the non-protective case}
\label{fig:np0}
\end{figure}

Again, we consider only those readouts for which $a_n$ measures 0. For the non-protective case, $n$ steps can be simulated using the circuit shown in figure \ref{fig:np0}.

\vspace{0.4cm}

In this case, the state of the quantum computer $\ket{\alpha'}$ after $n$ steps will be as follows.

\begin{align}
	\ket{\alpha'} &= \frac{\ket{\psi} + \ket{\psi}_{x-ng}}{2} \ket{0}_p  + \frac{\ket{\psi} - \ket{\psi}_{x-ng}}{2} \ket{1}_p
\end{align} 

If we consider the readouts for which the qubit $p$ measures 0 or 1, we obtain $P'_T(n)$ or $P'_R(n)$, respectively.

\section{Results and Discussion}
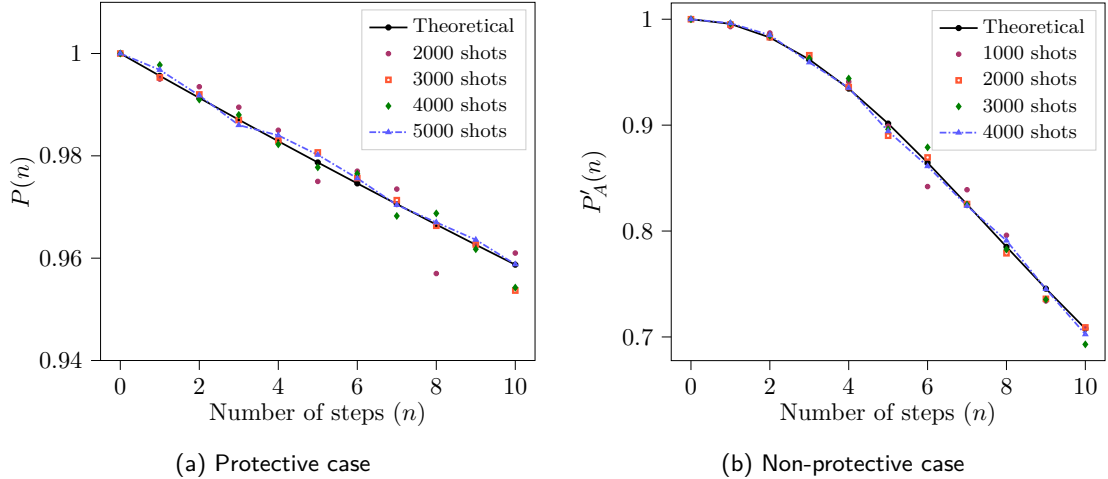
\begin{figure}[H]
\centering
\begin{subfigure}[b]{.5\textwidth}
  \centering
    \centering \resizebox{.95\linewidth}{!}{\begin{tikzpicture}

\definecolor{darkgray176}{RGB}{176,176,176}
\definecolor{green}{RGB}{0,128,0}
\definecolor{lightgray204}{RGB}{204,204,204}
\definecolor{orange}{RGB}{255,165,0}

\definecolor{crimson2143940}{RGB}{214,39,40}
\definecolor{darkgray176}{RGB}{176,176,176}
\definecolor{darkorange25512714}{RGB}{255,127,14}
\definecolor{forestgreen4416044}{RGB}{44,160,44}
\definecolor{lightgray204}{RGB}{204,204,204}
\definecolor{mediumpurple148103189}{RGB}{148,103,189}
\definecolor{steelblue31119180}{RGB}{31,119,180}

\definecolor{blue1}{RGB}{138, 138, 255}
\definecolor{blue2}{RGB}{92, 92, 255}
\definecolor{deepred}{RGB}{170, 51, 106}
\definecolor{orange}{RGB}{255, 87, 51}

\begin{axis}[
legend cell align={left},
legend style={fill opacity=1, draw opacity=1, text opacity=1, draw=lightgray204},
tick align=outside,
tick pos=left,
x grid style={darkgray176},
xlabel={Number of steps ($n$)},
xmin=-0.5, xmax=10.5,
xtick style={color=black},
y grid style={darkgray176},
ylabel={\(\displaystyle P(n)\)},
ymin=0.94, ymax=1.01,
ytick style={color=black}
]

\addplot [thick, black, mark=*, mark size=1, mark options={solid}]
table {%
0 1
1 0.995630145126226
2 0.991317244096027
3 0.987060066911328
4 0.982857420548247
5 0.978708147536386
6 0.974611124604331
7 0.970565261387916
8 0.966569499197769
9 0.962622809842958
10 0.958724194507729
};
\addlegendentry{\footnotesize{Theoretical}}

\addplot [line width=0.5pt, deepred, densely dashdotted, mark=*, mark size=1, mark options={solid}, only marks]
table {%
0 1
1 0.995
2 0.9935
3 0.9895
4 0.985
5 0.975
6 0.977
7 0.9735
8 0.957
9 0.963
10 0.961
};
\addlegendentry{\footnotesize{2000 shots}}
\addplot [line width=1pt, orange, densely dashdotted, mark=square, mark size=1, mark options={solid}, only marks]
table {%
0 1
1 0.995333333333333
2 0.992
3 0.987
4 0.983
5 0.980666666666667
6 0.975666666666667
7 0.971333333333333
8 0.966333333333333
9 0.962666666666667
10 0.953666666666667
};
\addlegendentry{\footnotesize{3000 shots}}
\addplot [line width=1pt, green, densely dashdotted, mark=diamond, mark size=1, mark options={solid}, only marks]
table {%
0 1
1 0.99775
2 0.991
3 0.988
4 0.98225
5 0.97775
6 0.9765
7 0.96825
8 0.96875
9 0.96175
10 0.95425
};
\addlegendentry{\footnotesize{4000 shots}}
\addplot [thick, blue2, densely dashdotted, mark=triangle, mark size=1, mark options={solid}]
table {%
0 1
1 0.9968
2 0.9918
3 0.986
4 0.984
5 0.9802
6 0.9756
7 0.9704
8 0.967
9 0.9636
10 0.9588
};
\addlegendentry{\footnotesize{5000 shots}}
\end{axis}

\end{tikzpicture}}
  \caption{Protective case}
  \label{fig:sub1}
\end{subfigure}%
\begin{subfigure}[b]{.5\textwidth}
  \centering
    \centering \resizebox{.95\linewidth}{!}{\begin{tikzpicture}

\definecolor{darkgray176}{RGB}{176,176,176}
\definecolor{green}{RGB}{0,128,0}
\definecolor{lightgray204}{RGB}{204,204,204}
\definecolor{orange}{RGB}{255,165,0}

\definecolor{crimson2143940}{RGB}{214,39,40}
\definecolor{darkgray176}{RGB}{176,176,176}
\definecolor{darkorange25512714}{RGB}{255,127,14}
\definecolor{forestgreen4416044}{RGB}{44,160,44}
\definecolor{lightgray204}{RGB}{204,204,204}
\definecolor{mediumpurple148103189}{RGB}{148,103,189}
\definecolor{steelblue31119180}{RGB}{31,119,180}

\definecolor{red1}{RGB}{252, 0, 0}
\definecolor{blue2}{RGB}{92, 92, 255}
\definecolor{deepred}{RGB}{170, 51, 106}
\definecolor{orange}{RGB}{255, 87, 51}

\begin{axis}[
legend cell align={left},
legend style={fill opacity=1, draw opacity=1, text opacity=1, draw=lightgray204},
tick align=outside,
tick pos=left,
x grid style={darkgray176},
xlabel={Number of steps ($n$)},
xmin=-0.5, xmax=10.5,
xtick style={color=black},
y grid style={darkgray176},
ylabel={\(\displaystyle P'_A(n)\)},
ymin=0.67765, ymax=1.01535,
ytick style={color=black}
]

\addplot [thick, black, mark=*, mark size=1, mark options={solid}]
table {%
0 1
1 0.995630145126226
2 0.982748395879233
3 0.962018518412417
4 0.934483556101105
5 0.901478897728775
6 0.864525168490726
7 0.825212876792357
8 0.785090906168803
9 0.745569492762732
10 0.707845621212714
};
\addlegendentry{\footnotesize{Theoretical}}

\addplot [line width=0.5pt, deepred, densely dashdotted, mark=*, mark size=1, mark options={solid}, only marks]
table {%
0 1
1 0.993
2 0.987
3 0.964
4 0.941
5 0.899
6 0.842
7 0.839
8 0.796
9 0.734
10 0.709
};
\addlegendentry{\footnotesize{1000 shots}}
\addplot [line width=1pt, orange, densely dashdotted, mark=square, mark size=1, mark options={solid}, only marks]
table {%
0 1
1 0.995
2 0.983
3 0.966
4 0.936
5 0.89
6 0.8695
7 0.8255
8 0.779
9 0.736
10 0.709
};
\addlegendentry{\footnotesize{2000 shots}}
\addplot [line width=1pt, green, densely dashdotted, mark=diamond, mark size=1, mark options={solid}, only marks]
table {%
0 1
1 0.996
2 0.984333333333333
3 0.963
4 0.944
5 0.896
6 0.879
7 0.825
8 0.782666666666667
9 0.735333333333333
10 0.693
};
\addlegendentry{\footnotesize{3000 shots}}
\addplot [thick, blue2, densely dashdotted, mark=triangle, mark size=1, mark options={solid}]
table {%
0 1
1 0.99625
2 0.985
3 0.95925
4 0.93525
5 0.89425
6 0.86125
7 0.82375
8 0.791
9 0.74525
10 0.7025
};
\addlegendentry{\footnotesize{4000 shots}}
\end{axis}

\end{tikzpicture}}
  \caption{Non-protective case}
  \label{fig:sub2}
\end{subfigure}
\caption{Each of the circuits in figure \ref{fig:circuit_protective} and \ref{fig:np0} were executed using noise free $\textit{statevector}$ simulator provided in Qiskit's $Aer$ simulator backend. The value of $\lambda$ was taken as 0.921.}
\label{fig:result1}
\end{figure}

\noindent
We first determine the constant $\lambda$ that appears in the phase shift implemented using the circuit shown in figure \ref{fig:shift}. We do this by initializing the qubits representing the pointer state with amplitudes of $\ket{\psi}_{x-g}$ using \ref{eq:10} and comparing it with those of $\ket{\psi}$ after transformation $U^{-1}_{\textnormal{QFT}}-\textit{phase shift}-U_{\textnormal{QFT}}$. We compare the state vectors in the two cases by varying $\lambda$ in the latter case to look for the maxima that occur for the same computational basis state in the two cases. We perform this simulation of translation of the pointer state using a noise-free $\textit{statevector}$ simulator provided in Qiskit's $Aer$ simulator backend. For $n=6$, $\lambda \in [0.913,0.926]$ and for $n=7$, $\lambda \in [0.921,0.926]$. For $n>7$, we do not obtain any better accuracy due to finite sampling of the output state vector \cite{Kreplin_2024}. Figure \ref{fig:result1} shows the results for the survival probability for the protective and non-protective case versus the number of steps when executed using $\textit{statevector}$ simulator with varying shots and the value of $\lambda$ taken as 0.921. 

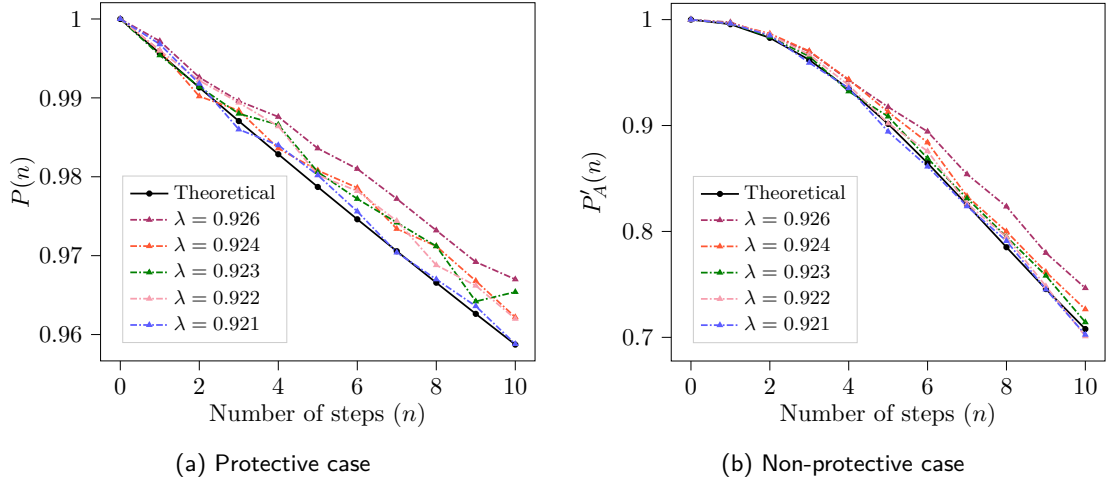
\begin{figure}[H]
\centering
\begin{subfigure}[b]{.5\textwidth}
    \centering \resizebox{.95\linewidth}{!}{\begin{tikzpicture}

\definecolor{darkgray176}{RGB}{176,176,176}
\definecolor{green}{RGB}{0,128,0}
\definecolor{lightgray204}{RGB}{204,204,204}
\definecolor{orange}{RGB}{255,165,0}

\definecolor{crimson2143940}{RGB}{214,39,40}
\definecolor{darkgray176}{RGB}{176,176,176}
\definecolor{darkorange25512714}{RGB}{255,127,14}
\definecolor{forestgreen4416044}{RGB}{44,160,44}
\definecolor{lightgray204}{RGB}{204,204,204}
\definecolor{mediumpurple148103189}{RGB}{148,103,189}
\definecolor{steelblue31119180}{RGB}{31,119,180}

\definecolor{red1}{RGB}{252, 0, 0}
\definecolor{blue2}{RGB}{92, 92, 255}
\definecolor{deepred}{RGB}{170, 51, 106}
\definecolor{orange}{RGB}{255, 87, 51}
\definecolor{pink0}{RGB}{247, 158, 174}

\begin{axis}[
legend style={at={(0.05,0.05)}, anchor=south west},
legend cell align={left},
legend style={fill opacity=1, draw opacity=1, text opacity=1, draw=lightgray204},
tick align=outside,
tick pos=left,
x grid style={darkgray176},
xlabel={Number of steps ($n$)},
xmin=-0.5, xmax=10.5,
xtick style={color=black},
y grid style={darkgray176},
ylabel={\(\displaystyle P(n)\)},
ymin=0.956660404233115, ymax=1.00206379027461,
ytick style={color=black}
]

\addplot [thick, black, mark=*, mark size=1, mark options={solid}]
table {%
0 1
1 0.995630145126226
2 0.991317244096027
3 0.987060066911328
4 0.982857420548247
5 0.978708147536386
6 0.974611124604331
7 0.970565261387916
8 0.966569499197769
9 0.962622809842958
10 0.958724194507729
};
\addlegendentry{\footnotesize{Theoretical}}

\addplot [thick, deepred, densely dashdotted, mark=triangle, mark size=1, mark options={solid}]
table {%
0 1
1 0.9972
2 0.9926
3 0.9896
4 0.9876
5 0.9836
6 0.981
7 0.9772
8 0.9732
9 0.9692
10 0.967
};
\addlegendentry{\footnotesize{$\lambda = 0.926$}}
\addplot [thick, orange, densely dashdotted, mark=triangle, mark size=1, mark options={solid}]
table {%
0 1
1 0.996
2 0.9902
3 0.9884
4 0.9836
5 0.9808
6 0.9786
7 0.9734
8 0.9712
9 0.9668
10 0.9622
};
\addlegendentry{\footnotesize{$\lambda = 0.924$}}
\addplot [thick, green, densely dashdotted, mark=triangle, mark size=1, mark options={solid}]
table {%
0 1
1 0.9954
2 0.9914
3 0.988
4 0.9866
5 0.9806
6 0.9772
7 0.9742
8 0.9712
9 0.9642
10 0.9654
};
\addlegendentry{\footnotesize{$\lambda = 0.923$}}
\addplot [thick, pink0, densely dashdotted, mark=triangle, mark size=1, mark options={solid}]
table {%
0 1
1 0.996
2 0.9922
3 0.9894
4 0.9864
5 0.9802
6 0.9782
7 0.9744
8 0.9688
9 0.9662
10 0.962
};
\addlegendentry{\footnotesize{$\lambda = 0.922$}}
\addplot [thick, blue2, densely dashdotted, mark=triangle, mark size=1, mark options={solid}]
table {%
0 1
1 0.9968
2 0.9918
3 0.986
4 0.984
5 0.9802
6 0.9756
7 0.9704
8 0.967
9 0.9636
10 0.9588
};
\addlegendentry{\footnotesize{$\lambda = 0.921$}}
\end{axis}

\end{tikzpicture}}
  \caption{Protective case}
  \label{fig:sub1}
\end{subfigure}%
\begin{subfigure}[b]{.5\textwidth}
  \centering \resizebox{.95\linewidth}{!}{\begin{tikzpicture}

\definecolor{darkgray176}{RGB}{176,176,176}
\definecolor{green}{RGB}{0,128,0}
\definecolor{lightgray204}{RGB}{204,204,204}
\definecolor{orange}{RGB}{255,165,0}

\definecolor{crimson2143940}{RGB}{214,39,40}
\definecolor{darkgray176}{RGB}{176,176,176}
\definecolor{darkorange25512714}{RGB}{255,127,14}
\definecolor{forestgreen4416044}{RGB}{44,160,44}
\definecolor{lightgray204}{RGB}{204,204,204}
\definecolor{mediumpurple148103189}{RGB}{148,103,189}
\definecolor{steelblue31119180}{RGB}{31,119,180}

\definecolor{red1}{RGB}{252, 0, 0}
\definecolor{blue2}{RGB}{92, 92, 255}
\definecolor{deepred}{RGB}{170, 51, 106}
\definecolor{orange}{RGB}{255, 87, 51}
\definecolor{pink0}{RGB}{247, 158, 174}

\begin{axis}[
legend style={at={(0.05,0.05)}, anchor=south west},
legend cell align={left},
legend style={fill opacity=1, draw opacity=1, text opacity=1, draw=lightgray204},
tick align=outside,
tick pos=left,
x grid style={darkgray176},
xlabel={Number of steps ($n$)},
xmin=-0.5, xmax=10.5,
xtick style={color=black},
y grid style={darkgray176},
ylabel={\(\displaystyle P'_A(n)\)},
ymin=0.67765, ymax=1.01535,
ytick style={color=black}
]

\addplot [thick, black, mark=*, mark size=1, mark options={solid}]
table {%
0 1
1 0.995630145126226
2 0.982748395879233
3 0.962018518412417
4 0.934483556101105
5 0.901478897728775
6 0.864525168490726
7 0.825212876792357
8 0.785090906168803
9 0.745569492762732
10 0.707845621212714
};
\addlegendentry{\footnotesize{Theoretical}}

\addplot [thick, deepred, densely dashdotted, mark=triangle, mark size=1, mark options={solid}]
table {%
0 1
1 0.9975
2 0.98475
3 0.97
4 0.94275
5 0.9175
6 0.89425
7 0.854
8 0.8235
9 0.77975
10 0.7465
};
\addlegendentry{\footnotesize{$\lambda = 0.926$}}
\addplot [thick, orange, densely dashdotted, mark=triangle, mark size=1, mark options={solid}]
table {%
0 1
1 0.9965
2 0.98675
3 0.97025
4 0.9435
5 0.91275
6 0.884
7 0.83325
8 0.8
9 0.76175
10 0.7265
};
\addlegendentry{\footnotesize{$\lambda = 0.924$}}
\addplot [thick, green, densely dashdotted, mark=triangle, mark size=1, mark options={solid}]
table {%
0 1
1 0.99625
2 0.98375
3 0.96525
4 0.93225
5 0.9085
6 0.869
7 0.83125
8 0.795
9 0.75825
10 0.71425
};
\addlegendentry{\footnotesize{$\lambda = 0.923$}}
\addplot [thick, pink0, densely dashdotted, mark=triangle, mark size=1, mark options={solid}]
table {%
0 1
1 0.9965
2 0.98575
3 0.967
4 0.938
5 0.902
6 0.87575
7 0.82625
8 0.79475
9 0.74825
10 0.701
};
\addlegendentry{\footnotesize{$\lambda = 0.922$}}
\addplot [thick, blue2, densely dashdotted, mark=triangle, mark size=1, mark options={solid}]
table {%
0 1
1 0.99625
2 0.985
3 0.95925
4 0.93525
5 0.89425
6 0.86125
7 0.82375
8 0.791
9 0.74525
10 0.7025
};
\addlegendentry{\footnotesize{$\lambda = 0.921$}}
\end{axis}

\end{tikzpicture}}
  \caption{Non-protective case}
  \label{fig:sub2}
\end{subfigure}
\caption{The circuits in figure \ref{fig:circuit_protective} and \ref{fig:np0} were executed using noise free $\textit{statevector}$ simulator by varying $\lambda$. The number of shots used for protective and non-protective case were 5000 and 4000, respectively.}
\label{fig:result2}
\end{figure}

Figure \ref{fig:result2} shows the same plots as in figure \ref{fig:result1}, except that here $\lambda$ is varied and the number of shots is fixed, 5000 and 4000 shots for the protective and non-protective case, respectively.

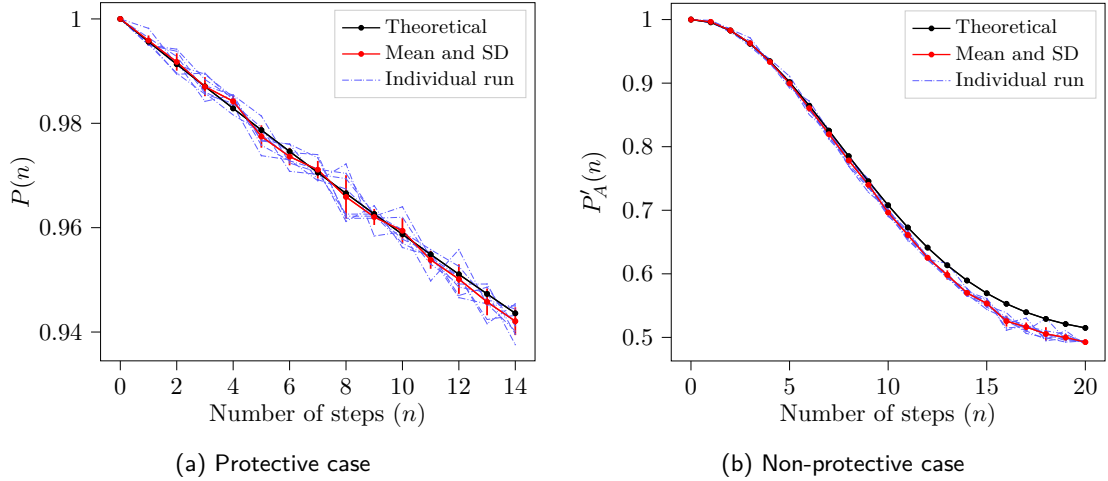
\begin{figure}[H]
\centering
\begin{subfigure}[b]{.5\textwidth}
  \centering \resizebox{.95\linewidth}{!}{\begin{tikzpicture}

\definecolor{crimson2143940}{RGB}{214,39,40}
\definecolor{darkgray176}{RGB}{176,176,176}
\definecolor{darkorange25512714}{RGB}{255,127,14}
\definecolor{darkturquoise23190207}{RGB}{23,190,207}
\definecolor{forestgreen4416044}{RGB}{44,160,44}
\definecolor{goldenrod18818934}{RGB}{188,189,34}
\definecolor{gray127}{RGB}{127,127,127}
\definecolor{lightgray204}{RGB}{204,204,204}
\definecolor{mediumpurple148103189}{RGB}{148,103,189}
\definecolor{orchid227119194}{RGB}{227,119,194}
\definecolor{sienna1408675}{RGB}{140,86,75}
\definecolor{steelblue31119180}{RGB}{31,119,180}
\definecolor{blue2}{RGB}{92, 92, 255}

\begin{axis}[
legend cell align={left},
legend style={fill opacity=1, draw opacity=1, text opacity=1, draw=lightgray204},
tick align=outside,
tick pos=left,
x grid style={darkgray176},
xlabel={Number of steps ($n$)},
xmin=-0.7, xmax=14.7,
xtick style={color=black},
y grid style={darkgray176},
ylabel={\(\displaystyle P(n)\)},
ymin=0.93448, ymax=1.00312,
ytick style={color=black}
]

\addplot [semithick, black, mark=*, mark size=1, mark options={solid}]
table {%
0 1
1 0.995630145126226
2 0.991317244096027
3 0.987060066911328
4 0.982857420548247
5 0.978708147536386
6 0.974611124604331
7 0.970565261387916
8 0.966569499197769
9 0.962622809842958
10 0.958724194507729
11 0.954872682678529
12 0.951067331118643
13 0.947307222887633
14 0.943591466405919
};
\addlegendentry{\footnotesize{Theoretical}}
\addplot [semithick, red, mark=*, mark size=1, mark options={solid}]
table {%
0 1
1 0.99585
2 0.99175
3 0.987025
4 0.984225
5 0.977475
6 0.973625
7 0.9711
8 0.9659
9 0.96205
10 0.9594
11 0.95385
12 0.95015
13 0.945775
14 0.942075
};
\path [draw=red, thick]
(axis cs:0,1)
--(axis cs:0,1);

\path [draw=red, thick]
(axis cs:1,0.994861314003336)
--(axis cs:1,0.996838685996664);

\path [draw=red, thick]
(axis cs:2,0.990089427809458)
--(axis cs:2,0.993410572190542);

\path [draw=red, thick]
(axis cs:3,0.985196236756712)
--(axis cs:3,0.988853763243288);

\path [draw=red, thick]
(axis cs:4,0.9830273460433)
--(axis cs:4,0.9854226539567);

\path [draw=red, thick]
(axis cs:5,0.975328054495335)
--(axis cs:5,0.979621945504665);

\path [draw=red, thick]
(axis cs:6,0.971950373175898)
--(axis cs:6,0.975299626824101);

\path [draw=red, thick]
(axis cs:7,0.969411805698387)
--(axis cs:7,0.972788194301613);

\path [draw=red, thick]
(axis cs:8,0.961747892101595)
--(axis cs:8,0.970052107898406);

\path [draw=red, thick]
(axis cs:9,0.960517844655396)
--(axis cs:9,0.963582155344605);

\path [draw=red, thick]
(axis cs:10,0.956987532383637)
--(axis cs:10,0.961812467616363);

\path [draw=red, thick]
(axis cs:11,0.952127356682305)
--(axis cs:11,0.955572643317695);

\path [draw=red, thick]
(axis cs:12,0.947313188409499)
--(axis cs:12,0.952986811590501);

\path [draw=red, thick]
(axis cs:13,0.943212896372119)
--(axis cs:13,0.948337103627881);

\path [draw=red, thick]
(axis cs:14,0.939412411973286)
--(axis cs:14,0.944737588026714);
\addlegendentry{\footnotesize{Mean and SD}}

\addplot [thin, blue2, densely dashdotted, mark=*, mark size=0.1, mark options={solid}]
table {%
0 1
1 0.9982
2 0.9912
3 0.9894
4 0.985
5 0.9758
6 0.9728
7 0.9712
8 0.9704
9 0.9624
10 0.957
11 0.9544
12 0.9486
13 0.9476
14 0.9376
};
\addlegendentry{\footnotesize{Individual run}}

\addplot [thin, blue2, densely dashdotted, mark=*, mark size=0.1, mark options={solid}]
table {%
0 1
1 0.995
2 0.9938
3 0.986
4 0.9834
5 0.9738
6 0.973
7 0.969
8 0.9722
9 0.9584
10 0.9592
11 0.9558
12 0.9526
13 0.9462
14 0.9444
};
\addplot [thin, blue2, densely dashdotted, mark=*, mark size=0.1, mark options={solid}]
table {%
0 1
1 0.995
2 0.9942
3 0.9884
4 0.9848
5 0.9768
6 0.976
7 0.9734
8 0.9622
9 0.9622
10 0.964
11 0.9542
12 0.9508
13 0.9416
14 0.9454
};
\addplot [thin, blue2, densely dashdotted, mark=*, mark size=0.1, mark options={solid}]
table {%
0 1
1 0.9958
2 0.9918
3 0.9842
4 0.9854
5 0.9814
6 0.9724
7 0.9706
8 0.9626
9 0.9626
10 0.9596
11 0.953
12 0.9492
13 0.9492
14 0.9396
};
\addplot [thin, blue2, densely dashdotted, mark=*, mark size=0.1, mark options={solid}]
table {%
0 1
1 0.9956
2 0.9894
3 0.9896
4 0.9844
5 0.9778
6 0.9736
7 0.9712
8 0.9612
9 0.9642
10 0.9562
11 0.9546
12 0.9466
13 0.9454
14 0.9402
};
\addplot [thin, blue2, densely dashdotted, mark=*, mark size=0.1, mark options={solid}]
table {%
0 1
1 0.9952
2 0.9896
3 0.9856
4 0.9838
5 0.9792
6 0.9708
7 0.9702
8 0.9694
9 0.9626
10 0.9578
11 0.9552
12 0.947
13 0.9486
14 0.9412
};
\addplot [thin, blue2, densely dashdotted, mark=*, mark size=0.1, mark options={solid}]
table {%
0 1
1 0.9956
2 0.9912
3 0.9872
4 0.9854
5 0.9766
6 0.976
7 0.9692
8 0.9674
9 0.9622
10 0.9594
11 0.9498
12 0.9558
13 0.9452
14 0.945
};
\addplot [thin, blue2, densely dashdotted, mark=, mark size=0.1, mark options={solid}]
table {%
0 1
1 0.9964
2 0.9928
3 0.9858
4 0.9816
5 0.9784
6 0.9744
7 0.974
8 0.9618
9 0.9618
10 0.962
11 0.9538
12 0.9506
13 0.9424
14 0.9432
};

\addplot [semithick, black, mark=*, mark size=1, mark options={solid}]
table {%
0 1
1 0.995630145126226
2 0.991317244096027
3 0.987060066911328
4 0.982857420548247
5 0.978708147536386
6 0.974611124604331
7 0.970565261387916
8 0.966569499197769
9 0.962622809842958
10 0.958724194507729
11 0.954872682678529
12 0.951067331118643
13 0.947307222887633
14 0.943591466405919
};
\addplot [semithick, red, mark=*, mark size=1, mark options={solid}]
table {%
0 1
1 0.99585
2 0.99175
3 0.987025
4 0.984225
5 0.977475
6 0.973625
7 0.9711
8 0.9659
9 0.96205
10 0.9594
11 0.95385
12 0.95015
13 0.945775
14 0.942075
};
\end{axis}

\end{tikzpicture}}
  \caption{Protective case}
  \label{fig:sub1}
\end{subfigure}%
\begin{subfigure}[b]{.5\textwidth}
  \centering \resizebox{.95\linewidth}{!}{\begin{tikzpicture}

\definecolor{crimson2143940}{RGB}{214,39,40}
\definecolor{darkgray176}{RGB}{176,176,176}
\definecolor{darkorange25512714}{RGB}{255,127,14}
\definecolor{darkturquoise23190207}{RGB}{23,190,207}
\definecolor{forestgreen4416044}{RGB}{44,160,44}
\definecolor{goldenrod18818934}{RGB}{188,189,34}
\definecolor{gray127}{RGB}{127,127,127}
\definecolor{lightgray204}{RGB}{204,204,204}
\definecolor{mediumpurple148103189}{RGB}{148,103,189}
\definecolor{orchid227119194}{RGB}{227,119,194}
\definecolor{sienna1408675}{RGB}{140,86,75}
\definecolor{steelblue31119180}{RGB}{31,119,180}
\definecolor{blue2}{RGB}{92, 92, 255}

\begin{axis}[
legend cell align={left},
legend style={fill opacity=1, draw opacity=1, text opacity=1, draw=lightgray204},
tick align=outside,
tick pos=left,
x grid style={darkgray176},
xlabel={Number of steps ($n$)},
xmin=-1, xmax=21,
xtick style={color=black},
y grid style={darkgray176},
ylabel={\(\displaystyle P'_A(n)\)},
ymin=0.462925, ymax=1.025575,
ytick style={color=black}
]

\addplot [semithick, black, mark=*, mark size=1, mark options={solid}]
table {%
0 1
1 0.995630145126226
2 0.982748395879233
3 0.962018518412417
4 0.934483556101105
5 0.901478897728775
6 0.864525168490726
7 0.825212876792357
8 0.785090906168803
9 0.745569492762732
10 0.707845621212714
11 0.672855325081405
12 0.64125377543634
13 0.613420820124419
14 0.589487200011542
15 0.569375235969446
16 0.552847377970744
17 0.539556497776407
18 0.529092941901619
19 0.521024847938566
20 0.514929783320556
};
\addlegendentry{\footnotesize{Theoretical}}
\addplot [semithick, red, mark=*, mark size=1, mark options={solid}]
table {%
0 1
1 0.99646875
2 0.98278125
3 0.96275
4 0.93353125
5 0.8996875
6 0.8608125
7 0.8198125
8 0.7784375
9 0.7400625
10 0.6965
11 0.66109375
12 0.62525
13 0.59815625
14 0.57003125
15 0.553375
16 0.5255625
17 0.516375
18 0.5054375
19 0.49984375
20 0.49253125
};
\path [draw=red, thick]
(axis cs:0,1)
--(axis cs:0,1);

\path [draw=red, thick]
(axis cs:1,0.995375893221872)
--(axis cs:1,0.997561606778128);

\path [draw=red, thick]
(axis cs:2,0.980887850692537)
--(axis cs:2,0.984674649307463);

\path [draw=red, thick]
(axis cs:3,0.958785312118212)
--(axis cs:3,0.966714687881788);

\path [draw=red, thick]
(axis cs:4,0.93092885476916)
--(axis cs:4,0.93613364523084);

\path [draw=red, thick]
(axis cs:5,0.8945264227142)
--(axis cs:5,0.9048485772858);

\path [draw=red, thick]
(axis cs:6,0.854867767562791)
--(axis cs:6,0.866757232437209);

\path [draw=red, thick]
(axis cs:7,0.814287344459203)
--(axis cs:7,0.825337655540797);

\path [draw=red, thick]
(axis cs:8,0.771926991667312)
--(axis cs:8,0.784948008332688);

\path [draw=red, thick]
(axis cs:9,0.733646273012899)
--(axis cs:9,0.746478726987101);

\path [draw=red, thick]
(axis cs:10,0.692667572570811)
--(axis cs:10,0.700332427429189);

\path [draw=red, thick]
(axis cs:11,0.65527092650641)
--(axis cs:11,0.66691657349359);

\path [draw=red, thick]
(axis cs:12,0.622627977879575)
--(axis cs:12,0.627872022120425);

\path [draw=red, thick]
(axis cs:13,0.590837676133631)
--(axis cs:13,0.605474823866369);

\path [draw=red, thick]
(axis cs:14,0.565553113005948)
--(axis cs:14,0.574509386994052);

\path [draw=red, thick]
(axis cs:15,0.547431253286016)
--(axis cs:15,0.559318746713984);

\path [draw=red, thick]
(axis cs:16,0.517016896232565)
--(axis cs:16,0.534108103767435);

\path [draw=red, thick]
(axis cs:17,0.509347153601565)
--(axis cs:17,0.523402846398435);

\path [draw=red, thick]
(axis cs:18,0.494865758480742)
--(axis cs:18,0.516009241519258);

\path [draw=red, thick]
(axis cs:19,0.494319743443387)
--(axis cs:19,0.505367756556613);

\path [draw=red, thick]
(axis cs:20,0.490487060444822)
--(axis cs:20,0.494575439555178);
\addlegendentry{\footnotesize{Mean and SD}}

\addplot [thin, blue2, densely dashdotted, mark=*, mark size=0.1, mark options={solid}]
table {%
0 1
1 0.99525
2 0.9845
3 0.961
4 0.93775
5 0.91075
6 0.86375
7 0.81825
8 0.7835
9 0.74175
10 0.69225
11 0.65625
12 0.627
13 0.594
14 0.57225
15 0.56125
16 0.51125
17 0.5195
18 0.5
19 0.50175
20 0.4935
};
\addlegendentry{\footnotesize{Individual run}}

\addplot [thin, blue2, densely dashdotted, mark=*, mark size=0.1, mark options={solid}]
table {%
0 1
1 0.99675
2 0.98075
3 0.9665
4 0.93325
5 0.9035
6 0.859
7 0.81675
8 0.7695
9 0.73525
10 0.697
11 0.66225
12 0.626
13 0.59525
14 0.56975
15 0.5575
16 0.51475
17 0.5175
18 0.51025
19 0.50525
20 0.49075
};
\addplot [thin, blue2, densely dashdotted, mark=*, mark size=0.1, mark options={solid}]
table {%
0 1
1 0.99625
2 0.97975
3 0.962
4 0.93525
5 0.89775
6 0.865
7 0.81325
8 0.7865
9 0.7425
10 0.6975
11 0.652
12 0.62925
13 0.59375
14 0.5665
15 0.56075
16 0.52575
17 0.51
18 0.53025
19 0.497
20 0.49375
};
\addplot [thin, blue2, densely dashdotted, mark=*, mark size=0.1, mark options={solid}]
table {%
0 1
1 0.997
2 0.9855
3 0.96075
4 0.93125
5 0.89925
6 0.859
7 0.8115
8 0.7785
9 0.74875
10 0.7015
11 0.654
12 0.622
13 0.59325
14 0.577
15 0.5475
16 0.5325
17 0.50625
18 0.4995
19 0.4925
20 0.4945
};
\addplot [thin, blue2, densely dashdotted, mark=*, mark size=0.1, mark options={solid}]
table {%
0 1
1 0.99875
2 0.98475
3 0.97125
4 0.93175
5 0.89225
6 0.87175
7 0.82475
8 0.76925
9 0.72825
10 0.70125
11 0.663
12 0.62725
13 0.597
14 0.568
15 0.544
16 0.5295
17 0.518
18 0.50575
19 0.49475
20 0.4945
};
\addplot [thin, blue2, densely dashdotted, mark=*, mark size=0.1, mark options={solid}]
table {%
0 1
1 0.995
2 0.9825
3 0.96225
4 0.937
5 0.89575
6 0.86125
7 0.8285
8 0.773
9 0.73725
10 0.69025
11 0.66675
12 0.62625
13 0.5945
14 0.56475
15 0.55575
16 0.5225
17 0.53025
18 0.49475
19 0.4985
20 0.491
};
\addplot [thin, blue2, densely dashdotted, mark=*, mark size=0.1, mark options={solid}]
table {%
0 1
1 0.99675
2 0.98275
3 0.96075
4 0.93075
5 0.8985
6 0.8565
7 0.824
8 0.78425
9 0.7485
10 0.6985
11 0.66825
12 0.6225
13 0.6165
14 0.5765
15 0.55
16 0.5295
17 0.51925
18 0.50625
19 0.49825
20 0.49375
};
\addplot [thin, blue2, densely dashdotted, mark=, mark size=0.1, mark options={solid}]
table {%
0 1
1 0.996
2 0.98175
3 0.9575
4 0.93125
5 0.89975
6 0.85025
7 0.8215
8 0.783
9 0.73825
10 0.69375
11 0.66625
12 0.62175
13 0.601
14 0.5655
15 0.55025
16 0.53875
17 0.51025
18 0.49675
19 0.51075
20 0.4885
};

\addplot [semithick, black, mark=*, mark size=1, mark options={solid}]
table {%
0 1
1 0.995630145126226
2 0.982748395879233
3 0.962018518412417
4 0.934483556101105
5 0.901478897728775
6 0.864525168490726
7 0.825212876792357
8 0.785090906168803
9 0.745569492762732
10 0.707845621212714
11 0.672855325081405
12 0.64125377543634
13 0.613420820124419
14 0.589487200011542
15 0.569375235969446
16 0.552847377970744
17 0.539556497776407
18 0.529092941901619
19 0.521024847938566
20 0.514929783320556
};
\addplot [semithick, red, mark=*, mark size=1, mark options={solid}]
table {%
0 1
1 0.99646875
2 0.98278125
3 0.96275
4 0.93353125
5 0.8996875
6 0.8608125
7 0.8198125
8 0.7784375
9 0.7400625
10 0.6965
11 0.66109375
12 0.62525
13 0.59815625
14 0.57003125
15 0.553375
16 0.5255625
17 0.516375
18 0.5054375
19 0.49984375
20 0.49253125
};
\end{axis}

\end{tikzpicture}}
  \caption{Non-protective case}
  \label{fig:sub2}
\end{subfigure}
\caption{The circuits in figure \ref{fig:circuit_protective} and \ref{fig:np0} were executed using default $Aer$ simulator with 5000 and 4000 shots for protective and non-protective case, respectively. The red line depicts the mean and standard deviation of eight data points for each step.}
\label{fig:result3}
\end{figure}

From the above figure it can be concluded that $\lambda = 0.921$ gives the optimal result for both cases. We use $\lambda = 0.921$ for all subsequent simulations. Figure \ref{fig:result3} shows the results when simulated using Qiskit's default $Aer$ simulator, depicting the mean and standard deviation for eight runs executed with 5000 and 4000 shots for the protective and non-protective case, respectively.

The deviations in results for different runs with the same number of shots can be attributed to the finite-sampling noise due to repeated circuit evaluations, even on error-free quantum computers. The deviations for all the steps are more in the protective case because of the larger number of gates and qubits used. We simulated the behaviour of photons undergoing decoherence (when they undergo weak interaction with the birefringent crystals), and the results match impressively with the theoretical predictions. The results of this shot-based noise-free simulation can be improved using a suitable regression technique.

\newpage
\appendix

\section{Experimental setup}
\noindent
We describe the relevant elements of the experimental setup as well as the operators associated with them. Let $\ket{H} = \begin{pmatrix} 1  \\ 0  \end{pmatrix}$ and $\ket{V} = \begin{pmatrix} 0  \\ 1  \end{pmatrix}$.  The operator for a HWP whose fast axis is inclined at an angle $\alpha$ is $W_{\alpha} = \begin{pmatrix} \cos 2\alpha & \sin 2\alpha \\ \sin 2\alpha & -\cos 2\alpha \end{pmatrix}$. In figure \ref{fig:setup}, the HWP b is placed before a PBS (not shown) and can be used to change the photon polarization to tune the intensity of the laser beam after PBS d. For HWPs e, g and h, the operators are $W_{\pi/8}$, $W_{3\pi/8}$ and $W_{\pi/4}$, respectively. All of the PBS transmit horizontally polarized photons and reflect vertically polarized ones. The direction of optic axis of the two birefringent crystals (calcite) i and j is as follows. For i, it is in the $xy-$plane and inclined $45^{\circ}$ with the $x-$axis and for j which is used only for phase compensation, it is along the $z-$axis. The operator of a polarizer oriented at an angle $\theta$ with the $x-$axis is $P_{\theta}=\ket{\theta}\bra{\theta}$, where $\ket{\theta} = \cos{\theta}\ket{H} + \sin{\theta}\ket{V}$. The operator for polarizer f is $P_{\pi/4}$.

\vspace*{4pt}   

\section{Wave function in position and momentum space}
\noindent
The wave function $\psi(x)$ of a free particle in the position space is related to its wave function in the momentum space $\phi(p)$ ($p$ is the momentum along the $x-$axis) as 

\begin{align}
	\phi (p) &= \frac{1}{\sqrt{2\pi \hbar}} \int_{-\infty}^{\infty} \psi(x) e^{-ipx/\hbar} dx = \mathcal{F} \big[ \psi(x) \big]
\end{align} 

\begin{align}
	\psi (x) &= \frac{1}{\sqrt{2\pi \hbar}} \int_{-\infty}^{\infty} \phi(p) e^{ipx/\hbar} dp = \mathcal{F}^{-1}  \big[\phi(p)\big]
 \label{eq:ft}
\end{align} 

where $\mathcal{F}$ and $\mathcal{F}^{-1}$ represent the Fourier transform and its inverse, respectively. Differentiating \ref{eq:ft},

\begin{align*}
	\frac{\partial \psi (x)}{\partial x} &= \frac{1}{\sqrt{2\pi \hbar}} \int_{-\infty}^{\infty} \frac{ip}{\hbar} \phi(p) e^{ipx/\hbar} dp \\
	\Rightarrow  \mathcal{F} \Bigg[ -i\hbar\frac{\partial \psi (x)}{\partial x} \Bigg] &= p \phi(p) \\
	\Rightarrow  \mathcal{F}^{-1} \mathcal{F} \Bigg[ -i\hbar\frac{\partial \psi (x)}{\partial x} \Bigg] &= \mathcal{F}^{-1} p \mathcal{F} \big[ \psi(x) \big] \\
	\Rightarrow -i\hbar\frac{\partial}{\partial x} &=  \mathcal{F}^{-1} p \mathcal{F} \\
	\Rightarrow \hat{p} &=  \mathcal{F}^{-1} p \mathcal{F}
\end{align*} 
It is important to note that the right side has to be applied to functions in momentum base in order for the equality to hold.
The relation \ref{eq:wf_qft} can also be obtained by using the shifting property of the Fourier transform. We have

\begin{align*}
	 \mathcal{F} \big[ \psi(x - g) \big] &= e^{-\frac{i}{\hbar}gp} \phi (p) \\
	 \Rightarrow  \psi(x - g) &= \mathcal{F}^{-1} e^{-\frac{i}{\hbar}gp}  \mathcal{F} \big[ \psi(x) \big] \\
	 \Rightarrow e^{-\frac{i}{\hbar}g \hat{p}} \psi (x) &=  \mathcal{F}^{-1} e^{-\frac{i}{\hbar}gp}  \mathcal{F} \big[ \psi(x) \big] \\
	 \Rightarrow e^{-\frac{i}{\hbar}g \hat{p}} \ket{\psi} &= U_{\textnormal{QFT}} e^{-\frac{i}{\hbar}g p} U^{-1}_{\textnormal{QFT}} \ket{\psi}
\end{align*} 

\comment{
Suppose we are given the values of $\psi(x)$ at certain values of $x$ that are discrete, say $\{x_0, x_1, ..., x_k, ..., x_{N-1}\}$ and evenly spaced, i.e., $\Delta x = x_{k+1} - x_k$. The approximate value of the Fourier integral in (14) can then be written as

\begin{align}
	\phi (p) &\approx \Delta x \sum_{k=0}^{N-1} e^{-ipx_k/\hbar}\psi(x_k)
\end{align}

The discrete values of $p$ to get $\phi(p)$ be the set of points $\{p_0, p_1, ..., p_j, ..., p_{N-1} \}$ are $p_j = \frac{2\pi \hbar j}{N\Delta x}$. Using $x_k = x_0 + k\Delta x$, we  have

\begin{align}
	\phi (p_j) &\approx \Delta x \sum_{k=0}^{N-1} e^{-\frac{i2\pi j}{N \Delta x} [ x_0 + k\Delta x ]}\psi(x_k) \nonumber \\
	&= \Delta x e^{-\frac{2\pi ij x_0}{N \Delta x}} \sum_{k=0}^{N-1} e^{-2\pi ijk / N} \psi(x_k) \nonumber \\
	&= \Delta x e^{-\frac{2\pi ij x_0}{N \Delta x}} \mathcal{F}_{\textnormal{DFT}} \{ \psi(x_k) \}_j
\end{align} 

where $\mathcal{F}_{\textnormal{DFT}} \{ \psi(x_k) \}_j$ denotes the $j^{th}$ term of the Discrete Fourier Transform (DFT) of $\psi(x_k)$.
}

Note that the Fourier transform $\mathcal{F}$ corresponds to the inverse quantum Fourier transform and viceversa, due to the fact that there is a $-$ sign in the argument for the exponential in $\mathcal{F}$ and a $+$ sign in that of $U_{QFT}$. 
Indeed, the quantum Fourier transform of the state given by \ref{eq:10} is $U_{QFT} \ket{\psi}$, where

\begin{align}
	U_{\textnormal{QFT}} &= \frac{1}{\sqrt N} \sum_{j,k=0}^{N-1} e^{2\pi ijk/N} \ket{k} \bra{j},
\end{align} 

so the inverse quantum Fourier transform $U^{-1}_{\textnormal{QFT}}$ is given by
 
 \begin{align}
	U^{-1}_{\textnormal{QFT}} = U_{QFT}^\dagger = \frac{1}{\sqrt N} \sum_{k,j=0}^{N-1} e^{-2\pi ijk/N} \ket{j} \bra{k}
\end{align}

\bibliography{references.bib}
\bibliographystyle{plain}
\end{document}